\documentclass[twocolumn,final,aps,prl,showpacs,superscriptaddress,amsmath,amssymb,amsfonts,floatfix]{revtex4-1}
\usepackage{graphicx}
\usepackage{multirow}
\usepackage{epsfig}

\newcommand{\mb}{\ensuremath{\mu_{\text{B}}}}

\def\be{\begin{equation}}
\def\ee{\end{equation}}

\usepackage{url}
\usepackage{color}
\usepackage[colorlinks,breaklinks,bookmarks=true,citecolor=blue,linkcolor=red,urlcolor=blue]{hyperref}
\definecolor{darkred}{rgb}{0.7,0.0,0}

\begin{document}

\title{Quantum Anomalous Hall State in Ferromagnetic SrRuO$_3$ (111) Bilayers}

\author{Liang Si}
\affiliation{Institut f\"{u}r Festk\"{o}rperphysik, TU Wien, Wiedner
Hauptstra{\ss}e 8-10, 1040 Vienna, Austria}

\author{Oleg Janson}
%\email{olegjanson@gmail.com}
\affiliation{Institut f\"{u}r Festk\"{o}rperphysik, TU Wien, Wiedner
Hauptstra{\ss}e 8-10, 1040 Vienna, Austria}

\author{Gang Li}
\email{gang.li@ifp.tuwien.ac.at}
\affiliation{Institut f\"{u}r Festk\"{o}rperphysik, TU Wien, Wiedner
Hauptstra{\ss}e 8-10, 1040 Vienna, Austria}
\affiliation{School of Physical Science and Technology, ShanghaiTech University, Shanghai 201210, China}

\author{Zhicheng Zhong}
%\email{Z.Zhong@fkf.mpg.de}
\affiliation{Max-Planck-Institut f\"{u}r Festk\"{o}rperforschung, Heisenbergstrasse 1, 70569 Stuttgart, Germany}

\author{Zhaoliang Liao}
\affiliation{MESA+ Institute for Nanotechnology, University of Twente, P.O. BOX 217, 7500AE, Enschede, The Netherlands}

\author{Gertjan Koster}
\affiliation{MESA+ Institute for Nanotechnology, University of Twente, P.O. BOX 217, 7500AE, Enschede, The Netherlands}

\author{Karsten Held}
\email{held@ifp.tuwien.ac.at}
\affiliation{Institut f\"{u}r Festk\"{o}rperphysik, TU Wien, Wiedner
Hauptstra{\ss}e 8-10, 1040 Vienna, Austria}

\date{\today}

\begin{abstract}
SrRuO$_3$ heterostructures grown in the (111) direction are a rare example of thin film ferromagnets. By means of density functional theory plus dynamical mean field theory we show that the half-metallic ferromagnetic state with an ordered magnetic moment of 2\,\mb/Ru survives the ultimate dimensional confinement down to a bilayer, even at elevated temperatures of 500$\,$K. In the minority channel, the spin-orbit coupling opens a gap at the linear band crossing corresponding to $\frac34$ filling of the $t_{2g}$ shell. We predict that the emergent state is Haldane's quantum anomalous Hall state with Chern number $C$=1, without an external magnetic field or magnetic impurities. 
\end{abstract}

%\pacs{73.20.At, 73.61.At, 73.43.Cd, 75.70.Cn}

\maketitle
The discovery of topological states of matter~\cite{TO:Wen,RevModPhys.82.3045,RevModPhys.83.1057} has significantly broadened and advanced our understanding of solid state physics. The historically first topological phenomenon is the quantum Hall effect~\cite{PhysRevLett.45.494} observed in a two-dimensional electronic system exposed to a strong external magnetic field. The quantum Hall effect manifests itself in the quantized transverse conductance (Hall conductance) stemming from nontrivial Berry curvatures of the filled Landau levels~\cite{Berry45,PhysRevLett.49.405,RevModPhys.82.1959} in an otherwise insulating state with a vanishing longitudinal conductance. As a result, a dissipationless edge mode appears along the boundary between the quantum Hall system and the vacuum.

In his Nobel prize winning work, Haldane~\cite{PhysRevLett.61.2015} realized that neither magnetic field nor Landau levels are actually required, but only breaking of time-reversal symmetry (TRS) and a non-trivial topology of the electronic structure. Arguably the easiest way to break TRS without magnetic field is through a spontaneous magnetization, inducing a quantized version of the conventional anomalous Hall effect, which is now known as the quantum anomalous Hall effect (QAH). Such insulating bulk systems are also called Chern insulators, as their topological edge states are characterized by the first Chern number, which is defined as an integral of the Berry curvature of the
filled bands over the Brillouin zone.

Recent research on quantum spin-Hall (QSH) insulators~\cite{ti:kane05,ti:koenig07,ti:hsieh08}, which can be viewed as two copies of a QAH state with equal but opposite magnetic moments, delivered an impressive number of candidate materials. In contrast, the search of QAH systems is far from being comprehensive owing to the strict conditions for realizing them, {\it i.e.}, realizing the spontaneous magnetization and the nontrivial Berry curvatures. One straightforward route is to magnetically dope a QSH insulator, where the presence of the magnetic impurity breaks the TRS and, thus, lifts the Kramers degeneracy of the QSH insulators to reach the QAH states~\cite{ti:chang13,ti:checkelsky14}. However, experimentally such extrinsic impurities are difficult to control, and need to generate a sufficiently strong magnetic moment while keeping the topology of the system unaffected. Thus, it is of more fundamental interest to have a genuine magnetic system that hosts the QAH state intrinsically.

In this context, heterostructures of transition-metal oxides (TMO) are a most promising material class. These artificial materials exhibit a plethora of interesting behaviors driven by electronic correlations and the dimensional confinement~\cite{interface:mannhart10,interface:rondinelli11,interface:zubko11,interface:chakhalian14}. Grown perpendicular to the body
diagonal of a perovskite lattice structure, i.e.\ along the $[111]$ direction, bilayers of perovskite TMO form a honeycomb lattice. The $d$-electrons in such bilayers can show a fascinating variety of electronic, magnetic, as well as topological~\cite{interface:xiao11,okamoto2014correlation} phases. Some heterostructures such as LaAuO$_3$-LaCrO$_3$ heterostructures were suggested as a realization of the QAH state~\cite{interface:liang13}. However, the fabrication of such heterostructures is at best challenging, because  bulk LaAuO$_3$ is not of the perovskite structure~\cite{interface:ralle93}. 
For another class of QAH candidate materials, freestanding monolayers of, e.g., NiCl$_3$ and OsCl$_3$ \cite{he2017near,PhysRevB.95.201402}, the necessity of an appropriate substrate is a serious limiting factor. Against this background, actual material realizations of Haldane’s
QAH state remain a great challenge.

In this letter, we investigate the electronic structure of (111)-oriented SrRuO$_3$ (SRO) bilayers sandwiched between SrTiO$_3$ (STO). This heterostructure can be routinely fabricated by advanced pulsed laser \cite{interface:grutter10} or metal-organic aerosol~\cite{interface:agrestini15} deposition, and thin SRO (111) films already exist and are ferromagentic \cite{grutter2010enhanced, PhysRevB.85.134429,PhysRevB.88.214410,ning2015anisotropy,ishigami2015thickness} \footnote{Magnetism in SRO films in general stems from the spin moment; x-ray magnetic circular dichroism shows only a small orbital moment \cite{grutter2010enhanced,ishigami2015thickness}.}. Using density functional theory (DFT) and dynamical mean field theory (DMFT) calculations, we show that SRO (111) bilayers remain ferromagnetic half-metals with a moment of 2\,\mb/Ru and a Curie temperature ($T_C$) exceeding room temperature. The band structure of the minority channel features a linear band crossing at $\frac34$ filling of the $t_{2g}$ shell, which becomes gapped due to the spin-orbit coupling (SOC). By simulating the respective microscopic Hamiltonian, we find a QAH state and the Chern number $C=1$. This result is further corroborated by a direct numerical solution of the full tight-binding Hamiltonian  parameterization of the Wannier functions on a long cylinder, which shows topological edge modes in the gap.

\begin{figure}[tb]
\includegraphics[width=8.6cm]{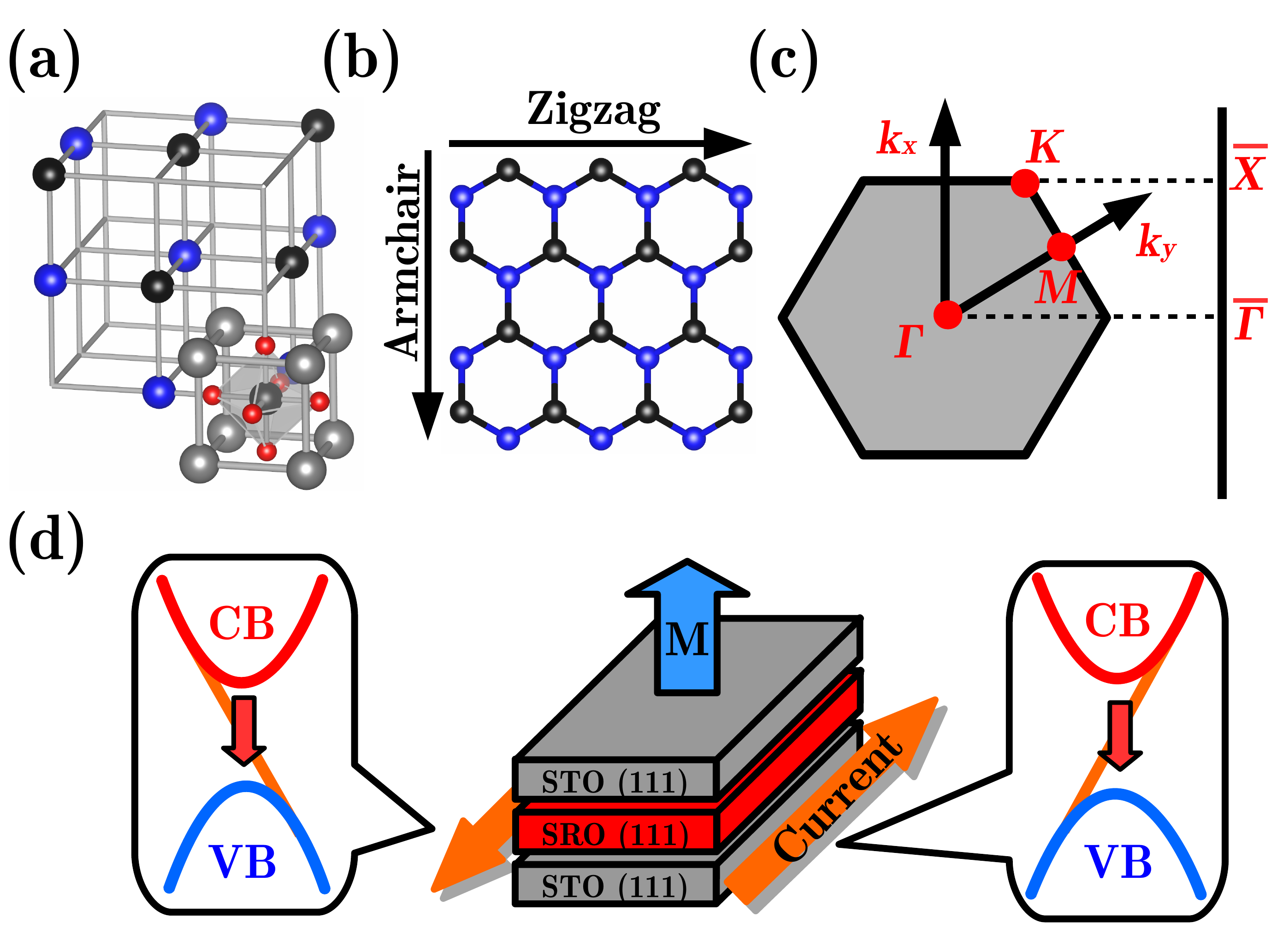}
\caption{(Color online) Crystal and electronic structure of 2\,SRO\,:\,7\,STO (111) superlattices. (a) Octahedral perovskite cage in
bulk SrRuO$_3$. Gray and red balls denote Sr and O atoms; blue and black ones Ru atoms in two honeycomb sublattices. (b) (111) bilayers of SRO form a honeycomb lattice. (c) High-symmetry points of the first Brillouin zone of the honeycomb lattice: $\Gamma$\,$(0,0,0)$, M\,$\left(0,\frac{\pi}{a},0\right)$ and K\,$\left(\frac{2\pi}{3a},\frac{2\pi}{3a},0\right)$. (d) Spin-polarized minority currents develop at the edge of the bilayer, and originate from the topological states between the valence band (VB) and the conduction band (CB) illustrated on the left and right, respectively.}
\label{Fig1}
\end{figure}

Before presenting the results of our DFT+DMFT calculations, we briefly explain how the unit cells are constructed. In the hypothetical cubic perovskite structure of SrRuO$_3$ [Fig.~\ref{Fig1}(a)], we consider layers stacked along the body diagonal $[111]$ of the cubic lattice. By cutting out two neighboring SRO monolayers [Fig.~\ref{Fig1}(b)], we obtain a bilayer whose Ru atoms form a honeycomb lattice. To comply with the periodic boundary conditions and to ensure a sufficient separation of individual SRO bilayers, we interleave them with seven layers of STO. The resulting 2\,SRO\,:\,7\,STO superlattice with 45 atoms in the unit cell has a pseudocubic structure described within the space group $P\bar{3}m1$ (164) with an inversion center in between the SRO monolayers. Please note that the Ru honeycomb in
SRO (111) is buckled, i.e., the blue and grey atoms in Fig.~\ref{Fig1}(b) are at a different (out of plane) height.

The smaller lattice constant of STO exerts a compressive strain of $\sim$0.45\,$\%$ on the SRO bilayers~\cite{koster2012structure}. To account for strain effects, we relax the $c$ unit cell parameter and the internal atomic coordinates within the generalized gradient approximation (GGA)~\cite{perdew1996generalized} as implemented in \textsc{vasp-5.12}~\cite{PhysRevB.48.13115, Kresse199615}. All further
electronic structure calculations were performed for this optimized structure using the \textsc{wien2k-14.2} \cite{blaha2001wien2k} code.

\begin{figure}[tb]
\includegraphics[width=8.6cm]{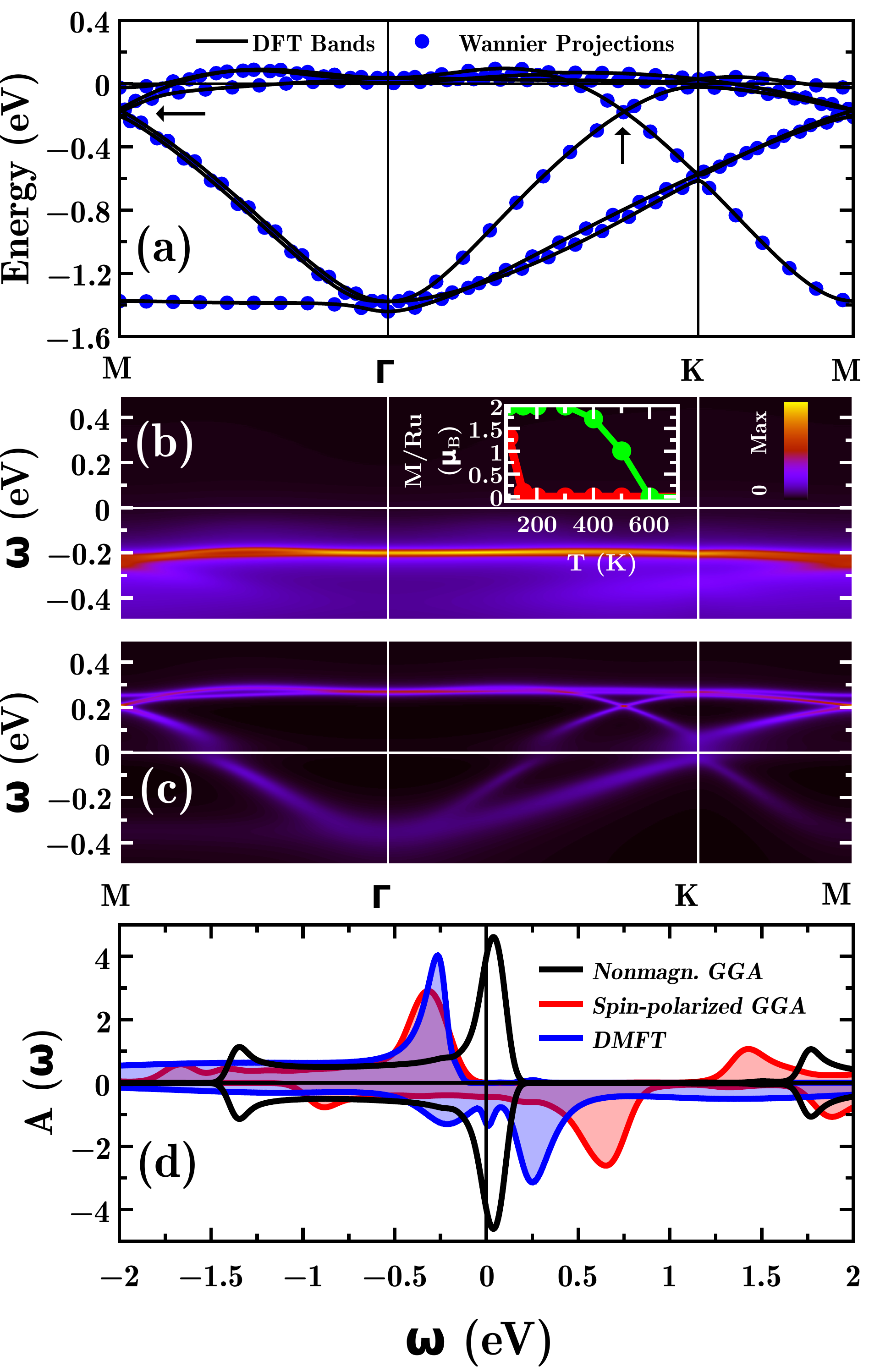}
\caption{(Color online) (a) Nonmagnetic scalar relativistic GGA band structure of 2\,SRO\,:\,7\,STO (solid lines) in comparison with the Fourier-transformed Wannier projections (circles). The arrow indicates the position of the Dirac point near $M$ point and along the $\Gamma$--K line. (b) spin-up and (c) spin-down DMFT spectral functions at 300\,K. (d) Density of states of nonmagnetic and magnetic GGA calculations in comparison with the DMFT spectral function. Inset in (b): ferromagnetic moment of bulk SRO (red) vs. the SRO (111) bilayer (green) as calculated by DFT+DMFT.}
\label{Fig2} 
\end{figure}

The nonmagnetic GGA band structure of 2\,SRO\,:\,7\,STO\,(111) is shown in Fig.~\ref{Fig2}(a). The manifold crossing the Fermi level comprises six Ru $t_{2g}$ bands. The higher-lying $e_g$ states (not shown) are empty and can be excluded from the correlated subspace. As a necessary preparatory step for DMFT calculations, we project the Ru $t_{2g}$ bands onto  maximally localized Wannier functions \cite{marzari2012maximally} using the \textsc{wien2wannier} code~\cite{kunevs2010wien2wannier}.

DMFT calculations were performed using the continuous time quantum Monte Carlo (CT-QMC) solver implemented in the code \textsc{w2dynamics}~\cite{parragh2012conserved}. The interaction parameters, the Coulomb repulsion $U$\,=\,3.0\,eV and the Hund's exchange $J$\,=\,0.3\,eV are adopted from~\cite{PhysRevB.92.041108}. We use the rotationally invariant form of the Hamiltonian, which implies an inter-orbital interaction $U'=U-2J$. For the analytical continuation of the resulting self-energy $\Sigma(i\omega_n)$ onto the real frequency axis $\omega$, the maximum entropy method~\cite{gubernatis1991quantum} was used.

DMFT reveals strong similarities between 2\,SRO\,:\,7\,STO (111) superlattices and bulk SrRuO$_3$: both are ferromagnetic and conducting
in the minority channel. The ferromagnetic half-metallic behavior is in sharp contrast to (001) SRO thin films that are insulators and lack
ferromagnetism~\cite{xia2009critical,PhysRevB.92.041108}. This difference stems from the spatial confinement which acts differently in (001) and (111) systems: Slicing in the $[001]$ direction lowers the onsite energy of the $xy$ orbital compared to the $xz$ and $yz$ orbitals~\cite{PhysRevB.92.041108}, favoring antiferromagnetism within the half-filled $xz$ and $yz$ orbitals. In contrast, the degeneracy of the $t_{2g}$ manifold is not affected by (111) slicing, and the on-site orbital energies are similar to bulk SrRuO$_3$. And yet, the spatial confinement is important: DMFT indicates that the ferromagnetic state of the 2\,SRO\,:\,7\,STO (111) superlattice survives up to $\sim$ 500\,K [inset of Fig.~\ref{Fig2}(b)], which is remarkably higher than the Curie temperature of bulk SRO (160\,K).
Since DMFT accurately reproduces the latter\cite{PhysRevB.85.134429}, this result is trustworthy, and it opens a route to long-sought-after ultra-thin layers that are ferromagnetic at room temperature \cite{note}.

Next, we evaluate the DMFT self-energy $\Sigma(\omega)$ on the real frequency axis and calculate the $k$-resolved spectral function $A(k,\omega)$ [Fig.~\ref{Fig2}(b) and \ref{Fig2}(c)]. The effect of the $k$-independent self-energy is twofold: the real part shifts the bands and renormalizes the band widths, while the imaginary part gives rise to broadening. The latter is particularly for the partially filled majority states: there is a broad incoherent continuum in Fig.~\ref{Fig2}(b) terminated by a nearly flat feature around 0.2\,eV below
the Fermi level. In contrast, the $k$-resolved spectral function for the minority channel  [Fig.~\ref{Fig2}(c)] largely resembles the GGA band structure [Fig.~\ref{Fig2}(a)]. The most prominent effect of electronic correlations here is the reduction of the band width down to $\sim$0.8\,eV, which is much smaller than that in the GGA (1.5\,eV) and in the bulk SRO (3.2\,eV). Altogether  the DMFT self energy essentially corresponds to a bandwidth quasiparticle renormalization of $Z\sim0.5$ and a reduction of the spin-splitting form 0.88$\,$eV in GGA to 0.47$\,$eV  in DFT+DMFT; otherwise, the electronic structure and its topology is not affected.

A remarkable property of $A(k,\omega)$ in the minority channel is the linear band crossing (at $\frac34$ filling of the $t_{2g}$ shell) near the $M$ point and along the $\Gamma$--K line. This feature inherited from the band structure [Fig.~\ref{Fig2}(a)] is resilient to electronic correlations [Fig.~\ref{Fig2}(c)]. But if the SOC is taken into account a gap opens at the band crossing: the full relativistic treatment within the modified Becke-Johnson potential~\cite{tran2009accurate} yields a tiny band gap of $\sim$1\,meV. Assuming the chemical potential is placed into the gap, the system becomes insulating at low temperatures. Topological properties of this envisaged insulating state are characterized by the Chern number $C$ of the occupied bands: zero or nonzero for a trivial state and the QAH state, respectively.

To determine $C$, we employ the full tight-binding Hamiltonian parameterization of the Wannier functions. The neighboring SRO bilayers are separated by seven SrTiO$_3$ layers; hence, we can restrict our analysis to a 2D model (couplings between the bilayers are well below
1 meV and hence can be neglected). The $|\phi_n(\vec{k})\rangle$ eigenvectors of the respective tight-binding Hamiltonian are calculated on a fine $k$-mesh of 120$\times$120 points and are used to calculate the Chern number $C$ \cite{ti:fukui07}:

\begin{equation}
C=\frac{1}{2\pi i}\sum_{n=1}^{N}\sum_{\vec{k}}ln U_x(\vec{k})U_y(\vec{k}+\triangle\vec{k}_x)U_x(\vec{k}+\triangle\vec{k}_y) U_y(\vec{k}),
\end{equation}

where $U_x= arg\langle \psi_n(\vec{k}) | \psi_n (\vec{k} + \triangle \vec{k}_x)\rangle$ are link variables \cite{ti:fukui07} connecting the neighbors in $\vec{k}$ mesh along $x$ direction  (analogously, for $U_y$ in the $y$ direction), and $N$ is the number of bands below the gap. In this way, we find $C$=1 for the bands below the SOC-induced gap; hence, the QAH state is realized.

\begin{figure*}[tb]
%\begin{minipage}{11.4cm}
\includegraphics[width=13.cm]{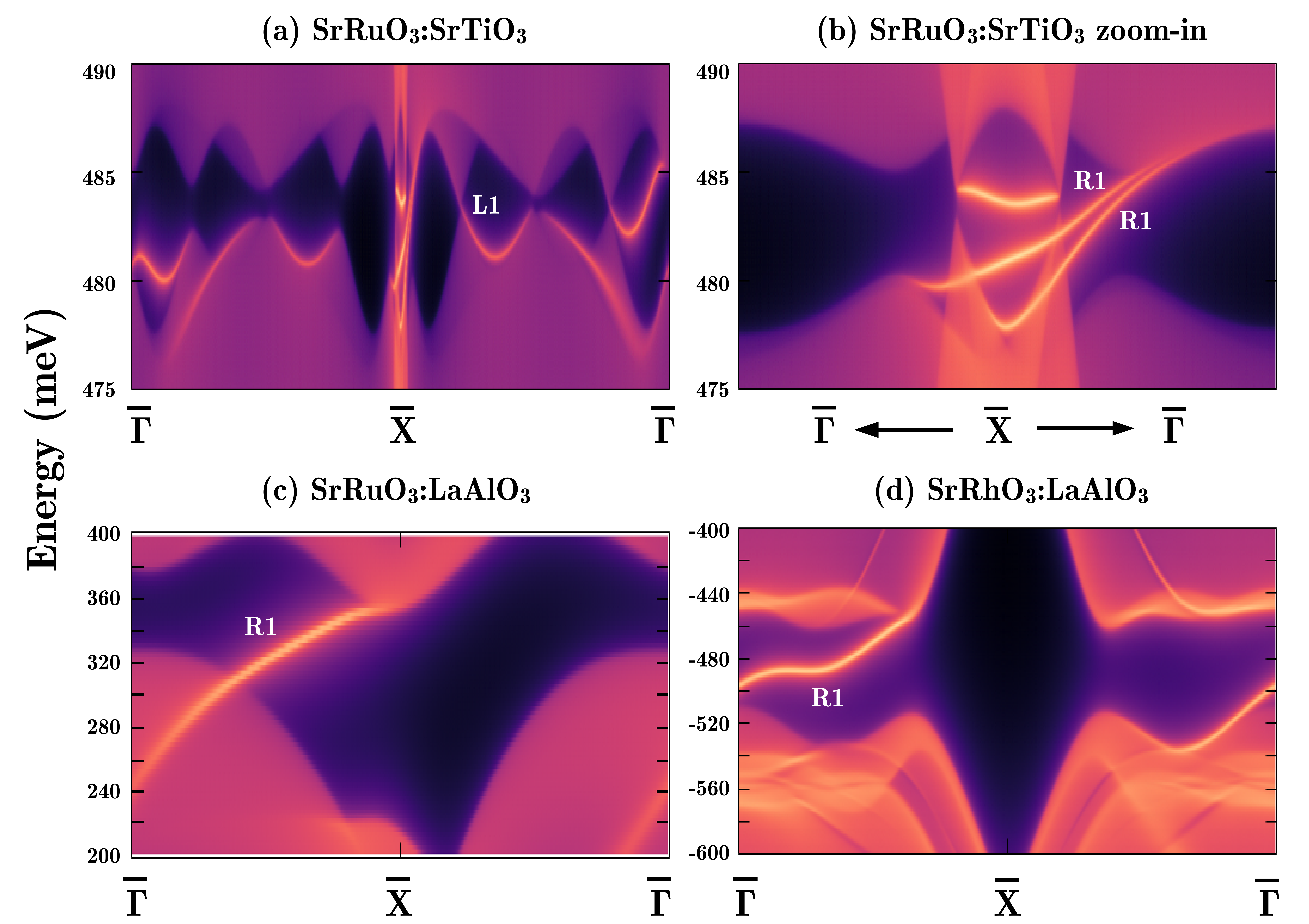}
%\end{minipage}\hfill
%\begin{minipage}{6.1cm}
\caption{opological edge states in the spectral functions $A(k_x,\omega)$ of (111) bilayers as calculated on semi-infinite cylinders. (a)
2SrRuO$_3$∶7SrTiO$_3$ along the full 1D $k_x$ path  $\bar{\Gamma} - \bar{X} - \bar{\Gamma}$  [as defined in Fig.~\ref{Fig1}(c)] and  (b) enlargement around $\bar{X}$, (c) 2SrRuO$_3$∶7LaAlO$_3$, and (d) 2SrRhO$_3$∶7LaAlO$_3$. Right (left) moving edge states are labeled with $R_1$ ($L_1$). The Chern number corresponds to the difference between the number of left- and right-moving edge states in the gap.}
\label{Fig3}
%\end{minipage}
\end{figure*}

To visualize this state, we recast our 2D Hamiltonian into mixed boundary conditions of a cylinder: periodic along one direction and open along the other. Spectral functions $A(k,\omega)$ are obtained using the iterative Green’s function approach \cite{sancho1985highly}, by successively doubling the cylinder length until the coupling between the two opposite edges becomes negligibly small, leading to an effectively semi-infinite cylinder. The resulting edge $A(k,\omega)$ [Fig.~\ref{Fig3}(a) and \ref{Fig3}(b)] reveals, besides the presence of trivial edge states that originate from the conduction (valence) band and rebound back to the conduction (valence) band in [Fig.~\ref{Fig3}(b)], also the topological edge states connecting the valence band and the conduction band. The number of “right moving” ($R_n$) and “left moving”($L_n$) states differ by unity, in accord with $C$=1 in our Chern number analysis. Let us also note that for the $C$=1 state, the consideration of the full microscopic model is needed: restricting the model to first-neighbor or first- and second-neighbor couplings only
yields unbalanced topological states as well, but with different Chern numbers.

For an experimental observation of edge states, the size of the SOC-induced band gap plays a crucial role. The gap size of $\sim$ 1meV in
2\,SRO∶7\,STO (111) and its position in the spectrum [$\sim$0.5\,eV above the chemical potential, see Fig.~\ref{Fig3}(a) and \ref{Fig3}(b)] confines the prospective QAH state to low temperatures and impedes its experimental observation. But oxide heterostructures are artificial materials with an inherent chemical flexibility, which is particularly advantageous for tuning the physical properties. Here, we explore the possibility of substitutional engineering to shift the band gap to lower energies and enhance its size.

A natural way of tuning a superlattice is the alteration of the substrate. Here, we replace SrTiO$_3$ with another commonly used substrate, LaAlO$_3$, and repeat our DFT-based analysis of Chern numbers as well as the momentum-resolved spectral function $A(k,\omega)$. In this way, we obtain a topological gap with $C$=1 above the Fermi level [Fig.~\ref{Fig3}(c)]. The larger gap of 30\,meV is very advantageous for a broad range of experimental techniques, as it mitigates the temperature and resolution constraints. For example, such a QAH state can be observed by angular resolved photoemission spectroscopy after laser pumping, as, e.g., demonstrated in Ref.~\cite{Phys.Rev.Lett.114.097401} for other topological surface states.

Further improvement can be achieved by substituting the transition metal atoms within the bilayer. In particular, replacing Ru with Rh changes the occupation of the $d$ level and shifts the topological gap below the Fermi level [Fig.~\ref{Fig3}(d)], so the resulting edge states can be detected in angular resolved photoemission spectroscopy.

In summary, we have shown that bilayers of SrRuO$_3$ on a SrTiO$_3$ substrate in the [111] direction emerge to be a candidate of long-sought-after room-temperature ferromagnetic half metals with an ordered moment of 2$\mu_B$ per Ru. The spin-orbit coupling opens a gap in the unoccupied part of the spectrum, which gives rise to a QAH state with the Chern number $C$=1. We show that the topological band gap
can be enhanced up to 30 meV and placed below the Fermi level by switching to a different substrate (LaAlO$_3$) and substitutional engineering at the interface (Rh instead of Ru), respectively.

\begin{acknowledgments}
L.\,S., G.\,L., O.\,J., and K.\,H. acknowledge financial support by European Research Council under the European Union’s Seventh Framework Program (FP7/2007–2013)/ERC through Grant Agreement No. 306447. L.\,S. also thanks the Austrian Science Fund (FWF) for support through
the Doctoral School W1243 Solids4Fun (Building Solids for Function). O.\,J. was supported by the FWF through the Lise Meitner program, Project No. M2050. Z.\,Z. acknowledges support by the FWF through the SFB ViCoM F4103. Z.\,L. acknowledges financial support from the European Union Council through the Seventh Framework Program (FP7), Grant No. NMP3-LA-2010-246102 IFOX. G.\,K. acknowledges funding from the DESCO program of the Dutch Foundation for Fundamental Research on Matter (FOM) with financial support from the Netherlands Organization for Scientific Research (NWO). We thank A.\,Sandvik for making available his maximum entropy program, and S.\,Okamoto for fruitful discussions. Calculations were done on the Vienna Scientific Cluster (VSC).
\end{acknowledgments}

\bibliographystyle{apsrev4-1}

\end{document}